\documentclass[runningheads]{llncs}
\usepackage[T1]{fontenc}
\usepackage{graphicx}
\usepackage{cite}
\usepackage{amsmath,amssymb,amsfonts}
\usepackage{enumitem}
\usepackage{algorithmic}
\usepackage[ruled,vlined]{algorithm2e}
\usepackage{textcomp}
\usepackage[numbers, sort]{natbib}
\usepackage{xcolor}
\usepackage{booktabs} 
\usepackage{pifont}       
\usepackage{makecell}
\usepackage{tabularx}
\usepackage{tikz}
\usetikzlibrary{shapes.geometric,arrows.meta,positioning}

\tikzset{
  block/.style = {rectangle, draw, fill=white!90, 
                  text width=10em, align=center, rounded corners, minimum height=3em},
  decision/.style = {diamond, draw, fill=white!90, 
                     text width=6em, align=center, aspect=2, inner sep=0pt},
  arrow/.style = {thick, >=Stealth}
}

\newcommand{\cmark}{\ding{51}}  
\newcommand{\xmark}{\ding{55}}  

\begin{document}

\author{Fernando Koch\inst{1} \and
Aladin Djuhera\inst{2}\and
Alecio Binotto \inst{3}}
\authorrunning{F. Koch et al.}
%
\institute{Florida Atlantic University, USA \\ \email{kochf@fau.edu} \and
Technical University Munich, Germany \\ \email{aladin.djuhera@tum.de} \and
Carl Zeiss AG, Germany \\ \email{alecio.binotto@zeiss.com} 
}

\newcommand{\CITE}{\textcolor{red}{CITE}}


\title{Intelligent Orchestration of Distributed Large Foundation Model Inference at the Edge}

\maketitle  


\begin{abstract}
    Large Foundation Models (LFMs), including multi-modal and generative models, promise to unlock new capabilities for next-generation Edge AI applications. However, performing inference with LFMs in resource-constrained and heterogeneous edge environments, such as Multi-access Edge Computing (MEC), presents significant challenges for workload orchestration due to time-varying network, compute, and storage conditions.
    In particular, current split inference strategies, which partition LFM layers across nodes, are not designed to adapt to fluctuating workloads, dynamic bandwidth conditions, or evolving privacy constraints in high-utilization MEC environments. 
    In this work, we propose a novel \emph{adaptive split inference orchestration framework} that elevates both the placement and partitioning of LFM layers to runtime-tunable variables. Specifically, our framework enables real-time, quality-of-service (QoS)-aware management of inference workloads by extending conventional orchestrators with three key services:
    (1) \emph{Capacity-aware workload distribution}, which continuously profiles node resources and selects an optimal subset of MEC nodes;
    (2) \emph{Dynamic partition migration}, which transparently relocates pre-cut LFM segments in response to changes in utilization or network conditions;
    (3) \emph{Real-time reconfiguration}, which dynamically re-splits LFM layers to balance latency, throughput, and privacy.
    We formalize the joint placement-partitioning problem, outline a reference architecture and algorithmic workflow, and discuss applicability in representative smart city, V2X, and industrial edge scenarios.
\end{abstract}


\section{Introduction}
\label{sec:introduction}

\textbf{Inference}, the real-time forward execution of a trained model, is typically less compute-intensive than model training and may require substantial resources for layer activation, memory, and model storage \cite{zhou2024survey}. This holds particularly for Large Foundation Models (LFMs) such as transformer-based large language models (LLMs) \cite{vaswani2017attention}, e.g., Llama \cite{grattafiori2024llama} and Qwen \cite{yang2025qwen3} models. Performing inference efficiently on the edge remains challenging, especially in Multi-Access Edge Computing (MEC) environments with limited and heterogeneous compute resources \cite{zhang2024beyond}.

\textbf{Distributed Split Inference} (DSI)~\cite{karjee2022split} has emerged as a promising approach to mitigate this problem. This strategy aims to partition an LFM into multiple segments that are executed sequentially across different nodes.

DSI strategies have typically employed \emph{static splits} of inference workloads, where some computation is executed locally, while heavier computation tasks are outsourced, alleviating the computational burden on the client device. Such \emph{splits} are mostly predetermined before execution and therefore lack adaptability to dynamic and heterogeneous operational conditions, such as fluctuating network reliability, changing node utilization, or intermittent connectivity. Consequently, these approaches produce suboptimal performance, compromising latency, resource utilization, and quality of service (QoS) guarantees, especially in mission-critical or latency-sensitive applications such as those found in financial services, industrial manufacturing, retail operations, and logistics \cite{karjee2021split}. This problem becomes even more acute in \emph{resource-constrained} and \emph{heterogeneous} edge environments, where multiple users rely on accessing shared edge compute resources, and where data privacy regulations (e.g., GDPR \cite{GDPR2016a}) often restrict offloading computations to the cloud. For instance, executing a 7B parameter LLM may incur 25~ms on an NVIDIA RTX A6000 edge node but over 250~ms on an NVIDIA Jetson accelerator. Meanwhile, backhaul latency can oscillate between sub-1 ms (mmWave) and 30 ms (congested Wi-Fi) \cite{struye2020towards, lau2023hybrid}. Such volatility renders any \emph{a priori} static split untenable.

Contemporary orchestration frameworks such as Kubernetes, Ray Serve, InferLine, and KubeEdge excel at container or micro-batch scheduling but treat AI models as black boxes, lacking mechanisms for runtime layer re-partitioning or privacy-aware placement \cite{vasireddy2023kubernetes}. 

Further, recent works, e.g., on jamming resilience for Split Federated Learning with LLMs\cite{djuhera2024rsfllmjammingresilientframework}, Federated Split Learning for satellite–terrestrial networks \cite{jiang2024federated}, and on Split Federated Mutual Learning (SFML) for traffic classification \cite{xia2025sfml}, address training-time collaboration, yet still assume fixed split points. Consequently, the key problem of \emph{joint, model-aware partition and placement} under dynamic edge conditions remains open.

In this work, we address this problem by introducing an \emph{adaptive split inference orchestration framework}, extending existing workload orchestration systems with domain-specific capabilities that are specifically tailored for LFMs, such as (multi-modal) LLMs. We introduce the following capabilities by leveraging the modular architecture of these models:

\begin{enumerate}
    \item \textbf{Distribution of workloads to edge nodes} that offer better performance or operational capacity than the original source node.
    
    \item \textbf{Redistribution of split LFM partitions} across connected edge nodes to dynamically optimize resources under changing conditions.
    
    \item \textbf{Adaptive reconfiguration of model splitting} (e.g., re-splitting) to further improve performance and resource utilization when required.
\end{enumerate}

Our framework operates on the \emph{computational graph of the LFM itself}, allowing decisions at the granularity of individual transformer blocks. This enables QoS-driven re-splitting that commodity orchestrators cannot express. Furthermore, since our solution emphasizes split inference, \emph{privacy} can be implemented as an additional feature at no cost if sensitive LFM layers can be executed locally, which makes reverse engineering data from model weights significantly more challenging for attackers \cite{xu2021privacy}. Through this approach, we establish a foundation for \emph{privacy-preserving, real-time}, and \emph{QoS-aware} AI inference in edge networks, aligning with key 6G objectives of seamless connectivity, low inference latency, and intelligent edge resource management \cite{letaief2021edge}. Importantly, our framework elevates privacy guarantees by localizing sensitive computations, thereby reducing regulatory exposure and aligning with emerging compliance standards.

Our contributions to the Edge AI landscape are multifold: 

\begin{itemize}
    \item a \textbf{reference architecture} for adaptive orchestration of distributed inference workloads;
    \item  a \textbf{method for dynamic redistribution} of model segments to accommodate fluctuating compute resources and connectivity, and;
    \item \textbf{establish mechanisms} for real-time, Service-Level Agreement (SLA)-compliant partitioning that align inference execution with QoS targets. 
\end{itemize}


\section{Background and Related Work}
\label{sec:background}

\subsection{Large Foundation Models at the Edge}

Next-generation 6G-enabled services, for example in dense urban environments, will need to support a multitude of AI-driven applications underpinned by LFMs \cite{saad2024artificialgeneralintelligenceaginative, llms_telecom}. However, deploying such large models typically requires significant computational resources and raises privacy concerns when handling sensitive data, making their adoption particularly challenging for inference in edge environments \cite{li2024llm, yao2024survey}. Further, as organizations strive to keep data on-premise (e.g., for regulatory compliance such as GDPR~\cite{GDPR2016a}), optimizing \emph{distributed inference} becomes a critical enabler of low-latency and privacy-preserving AI services~\cite{li2019edge, zhou2024survey}. 

For that, next-generation networks make use of MEC infrastructures \cite{abderrahime2020mec}, which embed computing resources directly within the network. While this brings compute closer to end-users, a single MEC-enabled base station can quickly become saturated by various inference workloads such as from smart city and crowd management, personalized user applications, or industrial applications \cite{lin2023pushing}. Looking ahead, 6G is envisaged to evolve beyond a mere network infrastructure upgrade into a \emph{trustworthy} \cite{fettweis20226g} and \emph{intelligent workload orchestration system}, enabling distributed, LFM-based AI services that seamlessly shift computation across user devices, edge, and cloud as network and compute conditions change \cite{camelo2022requirements, zeb2023toward}. 

\subsection{Challenges in AI Workload Orchestration}

The current industry norm has been to integrate general-purpose orchestration platforms (e.g., Kubernetes or proprietary MEC orchestrators), which facilitate application deployment and scaling but were not designed for challenges in inference scaling of modern LFM architectures, including fine-grained model partitioning and real-time, potentially hardware-accelerated inference optimization~\cite{chen2021enhancing}. Existing solutions thus primarily target stateless services or relatively simple microservices, neglecting the unique requirements of edge-based AI pipelines, particularly for dynamic model splitting, QoS-driven scheduling, and adaptive resource reallocation \cite{chen2024adaptive}. 

In particular, in Edge AI scenarios involving LFMs, traditional workload orchestration platforms fall short of meeting Edge AI-centric inference demands as they lack mechanisms to dynamically redistribute or reconfigure large model partitions based on real-time changes in \emph{network conditions, node utilization}, or \emph{connectivity} \cite{carrion2022kubernetes}. While model-serving stacks such as Triton, InferLine, Ray~Serve, and MLC-Serve introduce batching or replica autoscaling, they still treat neural networks as opaque binaries and cannot \emph{re-partition} a model graph at runtime.  Edge extensions like KubeEdge and OpenYurt inherit the same pod-level abstraction, leaving the joint problem of \emph{runtime layer splitting and placement}, especially under privacy and QoS constraints, unaddressed (see Table~\ref{tab:orc_gap}).

\begin{table*}[t]
\centering
\scriptsize
\caption{Comparison of \textbf{capability coverage} across mainstream workload orchestrators / serving stacks and our proposed \emph{Adaptive Split Orchestrator}.  
\cmark\ indicates native support, \xmark\ indicates the capability is absent, and entries in parentheses denote the mechanism (e.g., \emph{pods}, \emph{replicas}).  
Our framework simultaneously (i) reasons over the \emph{layer graph}, (ii) adapts \emph{split points} and \emph{placements} \textit{at runtime}, (iii) schedules under explicit \emph{QoS/SLA} targets, and (iv) enforces \emph{privacy‐aware} layer placement.}

\label{tab:orc_gap}
\begin{tabular}{lccccc}
\toprule
\textbf{Orchestrator}
 & \textbf{\makecell{Layer‐Aware\\Graph}}
 & \textbf{\makecell{Runtime\\Split}}
 & \textbf{\makecell{Placement\\Adapt.}}
 & \textbf{\makecell{QoS‐\\Aware}}
 & \textbf{\makecell{Privacy\\Support}} \\
\midrule
Kubernetes                 & \xmark & \xmark & \makecell{\cmark\\(pods)}                & \xmark           & \xmark \\
KubeEdge                   & \xmark & \xmark & \makecell{\cmark\\(edge pods)}           & \xmark           & \xmark \\
Ray Serve                  & \xmark & \xmark & \makecell{\cmark\\(replicas)}            & \makecell{\cmark\\(latency tiers)} & \xmark \\
NVIDIA Triton              & \xmark & \xmark & \makecell{\cmark\\(inst.\ groups)}       & \makecell{\cmark\\(batch/queue)}   & \xmark \\
InferLine                  & \xmark & \xmark & \makecell{\cmark\\(replicas)}            & \makecell{\cmark\\(tail‐latency)} & \xmark \\
\midrule
EdgeShard~\cite{EdgeShard}  & \makecell{\cmark\\(manual)} & \makecell{\cmark\\(manual)} & \xmark & \xmark & N/A \\
\textbf{Ours}              & \cmark & \makecell{\cmark\\(runtime)} & \makecell{\cmark\\(runtime)} & \makecell{\cmark\\(SLA‐aware)} & \makecell{\cmark\\(layer scope)} \\
\bottomrule
\end{tabular}
\end{table*}

\ 

As a result:
\begin{itemize}
    \item \textbf{Latency spikes} occur when critical links become congested, delaying real-time applications.
    \item \textbf{Straggler problems} arise when tasks are bottlenecked on overloaded or slower nodes, degrading overall QoS.
    \item \textbf{Resource utilization} becomes imbalanced, either overloading certain nodes or leaving others underutilized, leading to missed SLAs.
    \item \textbf{Privacy risks} escalate when large volumes of sensitive data must be offloaded to remote servers due to inadequate local processing.
\end{itemize}

While current research predominantly focuses on efficient AI model \emph{training} (e.g., hardware-efficient training \cite{ duan2024efficienttraininglargelanguage}, quantization \cite{lang2024comprehensive}, and federated learning \cite{yao2024federated}), the practical challenges of \emph{inference scaling and efficiency} at the edge remain relatively overlooked \cite{zhou2024survey}. Yet, these inference-related challenges are increasingly critical for the widespread adoption of LFMs in industrial and commercial scenarios, particularly in future \emph{AI as a Service} (AIaaS)-driven 6G networks \cite{saad2024artificialgeneralintelligenceaginative}.

\subsection{Distributed and Adaptive Split Inference}

There are different techniques to distribute workloads on heterogeneous computing environments. For example, the authors of \cite{alecio1, alecio2} proposed heuristics to distribute workloads among CPUs and GPUs (or any computing units), monitoring their execution and dynamically adapting the overall scheduling giving changing conditions to, e.g., maximize performance or minimize energy consumption. 

\ 

To alleviate computational demands particularly present in inference workloads, \emph{distributed split inference} (DSI) \cite{mohammed2020distributed} has emerged as an approach within Edge AI, which partitions a model across different compute locations (e.g., client device, MEC node, cloud) to balance local processing with remote offloading. Here, early, lightweight, or privacy-sensitive model layers are often chosen to be executed locally on-device or on trusted MEC nodes, extracting compact feature maps to reduce data transmission overhead. Subsequent layers are then offloaded to remote cloud environments and processed by more powerful servers \cite{mohammed2020distributed, timor2025distributed}.

\ 

Although DSI enables larger AI models to operate closer to data sources, current implementations predominantly employ static splits defined \textit{a priori} based on expected conditions, without runtime adaptation. While some studies, such as EdgeShard \cite{EdgeShard}, explore collaborative inference setups where a model is shared across edge nodes, these approaches continue to lack dynamic orchestration of model splits and thus cannot effectively respond to real-world changes in edge environments, such as fluctuating node workloads, intermittent connectivity, variable network reliability, or dynamically changing service demands. As a result, traditional solutions frequently lead to suboptimal latency, inefficient resource utilization, degraded service quality, and decreased compliance with QoS guarantees and SLAs \cite{hudson2024qos}. Consequently, recent research highlights the benefits of \emph{adaptive split inference}, wherein partition points or even partition strategies (e.g., layer reordering) can be reconfigured at runtime to maintain QoS under shifting conditions \cite{chen2024adaptive}. This approach, combined with optimal orchestration policies, has the potential to cater to the increasingly demanding AI inference workloads in future AIaaS 6G-enabled networks and edge environments. 

\newpage

In addition to inference, recent split learning approaches illustrate the state of today’s limitations from the model training point of view. 
R-SFLLM \cite{djuhera2024rsfllmjammingresilientframework} freezes an LLM split immediately after the embedding layer of a transformer so that raw tokens never leave the device, and then analyses jamming resilience for that single layout.
FedSL-LSTM \cite{jiang2024federated} for satellite–terrestrial anomaly detection pre-divides each LSTM into client- and server-side subnetworks before training begins. The interface remains unchanged to keep gateway compute low and ground timing predictable.
SFML \cite{xia2025sfml} adopts a fixed three-segment CNN in which only lightweight ``head+tail” layers stay on edge routers while the compute-heavy core runs in the cloud, simplifying privacy guarantees but eliminating runtime flexibility.
Across all three studies, split points are determined \emph{a priori} and never migrated, leaving questions of dynamic re-splitting, QoS-driven placement, and privacy-aware adaptation unanswered, precisely the gaps our adaptive orchestrator addresses.
Table~\ref{tab:split_related} summarizes these representative frameworks, highlighting their fixed partitioning schemes and the lack of adaptive capabilities.

\begin{table}[t]
\centering
\scriptsize
\caption{Comparison of representative split learning frameworks with respect to their application context, model partitioning strategies, and (lack of) adaptivity. While each approach adopts fixed split points to simplify deployment or privacy guarantees, none supports dynamic re-partitioning or runtime migration, highlighting the need for adaptive orchestration in real-world, QoS-sensitive Edge AI scenarios.}
\label{tab:split_related}
\begin{tabularx}{\linewidth}{l*{3}{>{\raggedright\arraybackslash}X}}
\toprule
 & \textbf{R-SFLLM} 
 & \textbf{FedSL-LSTM} 
 & \textbf{SFML} \\
\midrule
\textbf{Application} 
 & Jamming-resilient FL with LLMs over 6G wireless links 
 & LSTM anomaly detection in satellite–terrestrial integrated networks 
 & Encrypted-traffic classification on edge routers + cloud \\

\addlinespace
\textbf{Split Layer} 
 & After the \emph{embedding} block; attention layers off-device 
 & First LSTM layers on gateways; remainder on ground server 
 & Lightweight head + tail on router; CNN core in cloud \\

\addlinespace
\textbf{Adaptivity} 
 & Static (single two-cut split, never moved during training)
 & Static (single two-cut split, never moved during training)
 & Static (single three-cut split, never moved during training) \\
\bottomrule
\end{tabularx}
\end{table}

\ 

Thus, in standard implementations, orchestrators cannot dynamically decide to offload additional LFM layers to another edge node or re-split the network, leaving significant performance and reliability gains unrealized. Note that this problem \emph{does not} originate from within the network itself as current 5G and future 6G architectures natively implement adaptive resource management strategies for various services and network slices \cite{thantharate2023adaptive6g}. In current deployments, effectiveness thus remains limited by the absence of real-time dynamic orchestration policies tailored explicitly to modern AI workloads and complex LFM deployment scenarios.
This is especially problematic for heterogeneous compute nodes, which complicates uniform deployment strategies. Thus, a \emph{one-size-fits-all} static partitioning rarely works, as local workloads, performance constraints, and available resources differ from one site to another \cite{li2024llm}.

\subsection{Key Design Goals for Adaptive LFM Split Inference}

Despite evidence that splitting models can significantly improve efficiency and privacy, practical deployments remain constrained by \emph{static} or \emph{coarse-grained} orchestration mechanisms~\cite{karjee2021split, zhou2019distributing}. Today’s solutions thus seldom adapt to shifting network or compute conditions in real time, leading to latency spikes, resource imbalances, and potential SLA and QoS violations~\cite{li2019learning}. Meanwhile, next-generation 6G network architectures will further exacerbate the complexity of distributing large-scale inference workloads across heterogeneous edge topologies to support various commercial and operational AIaaS applications~\cite{llms_telecom}. 

\ 

\noindent
Hence, an \emph{adaptive split inference} framework will be required that:

\begin{enumerate}
    \item dynamically reconfigures the partition of LFM layers among edge and cloud compute nodes,
    \item exploits real-time profiling of resource availability and network conditions,
    \item preserves data privacy by keeping sensitive computations locally, and
    \item ensures consistent, QoS-compliant performance under fluctuating workloads.
\end{enumerate}

\noindent
In the next section, we propose a novel orchestration method that closes this gap by intelligently managing LFM inference across edge compute infrastructures.

\section{Proposal}
\label{sec:proposal}

Recall the dense urban environment scenario from earlier where multiple users split their inference workloads between their local devices and an edge node (e.g., 5G-MEC). To this end, the LFMs must be partitioned accordingly, resulting in several different split configurations depending on local compute capacity, privacy requirements, and edge node capacity. In this scenario, we need to address the following three problems:
\begin{enumerate}
    \item How to ensure that split inference \emph{can indeed} take place on the assigned edge node given QoS and/or SLA requirements?
    \item How to \emph{redistribute} the split inference request to other candidate nodes in a connected region in case inference on the originally assigned node is not possible?
    \item How to \emph{dynamically revise} suboptimal LFM splits to obtain the best possible configuration given the local- and wide-area edge compute capacity?
\end{enumerate}

\begin{figure*}[h!]
    \centering
    \includegraphics[width=0.9\textwidth]{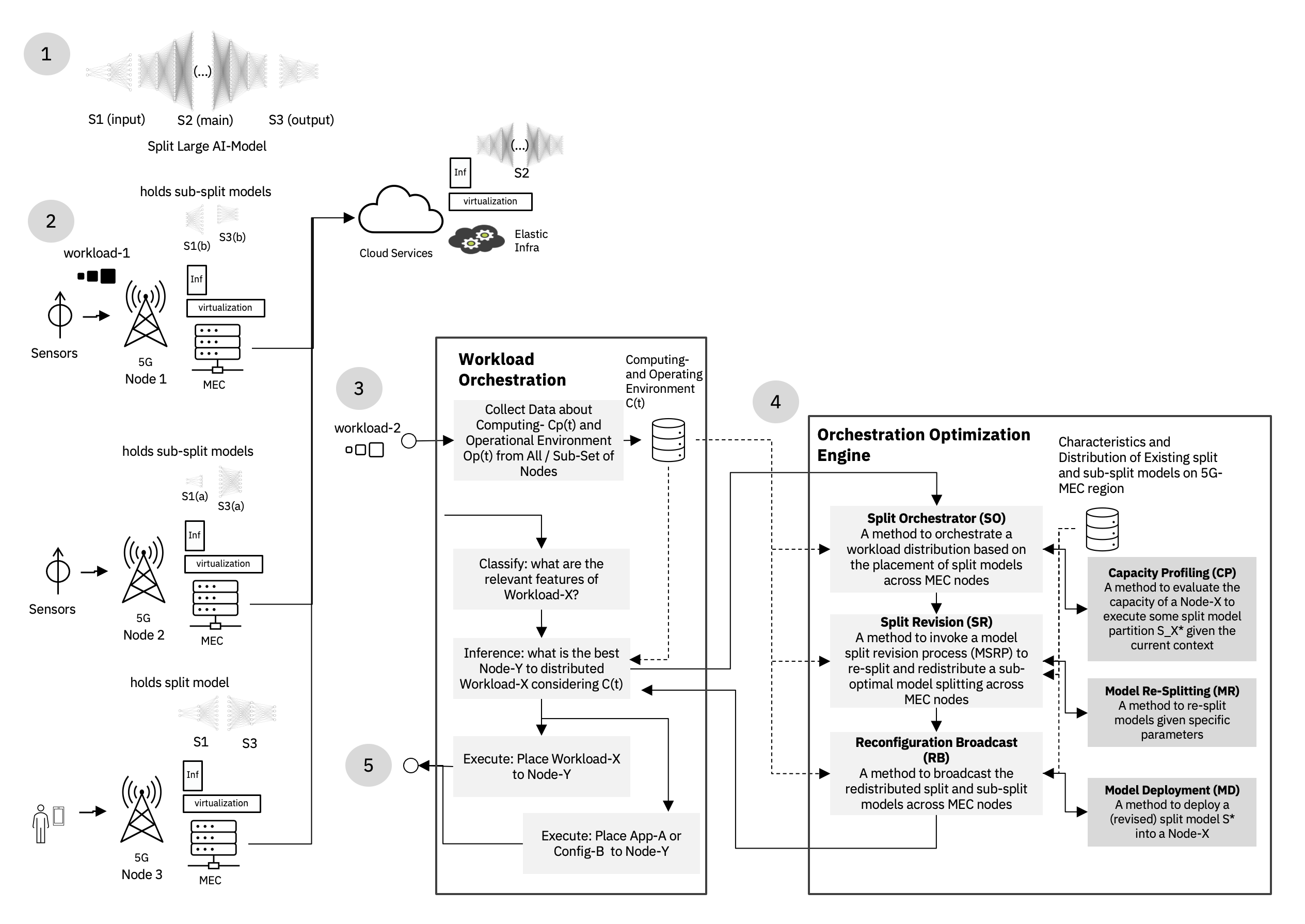}
    \caption{Reference architecture of the proposed adaptive split inference orchestration. Sub-split models (S1, S2, S3) are deployed across edge/cloud nodes, while a central orchestrator, guided by real-time capacity profiling, re-splits and reconfigures workloads on demand to meet QoS and privacy constraints. A corresponding workflow diagram of our proposed Algorithm \ref{alg:workflow} is given in Figure \ref{fig:algorithm_workflow}.}
    \label{fig:overview}
\end{figure*}

To address these challenges, we propose an \emph{adaptive split inference orchestration} framework that dynamically manages LFM partitions across heterogeneous edge nodes. Figure~\ref{fig:overview} depicts a possible realization of this framework in a 5G/6G-MEC deployment, including key components for monitoring, decision-making, model partitioning, and reconfiguration. We outline a detailed reference architecture as follows.

\subsection{Reference Architecture}

Our framework orchestrates on-demand allocation and reallocation of LFM partitions under evolving operational conditions via the following core modules:

\begin{enumerate}
    \item \textbf{Monitoring \& Capacity Profiling (CP):}
    Collects real-time metrics from edge nodes and the network environment, such as CPU/GPU utilization, memory usage, bandwidth, and latency. These metrics guide the orchestrator in partition placement and potential re-splitting decisions. \\

    \item \textbf{Adaptive Orchestrator (AO):}
    Acts as the decision-making engine by evaluating whether to:
    \begin{itemize}
        \item \emph{Keep} the current split (no changes).
        \item \emph{Redistribute} sub-splits across underutilized or more capable nodes.
        \item \emph{Fully re-split} the model to find an updated partition configuration.
    \end{itemize}
    These decisions are informed by constraints like node capacity, privacy requirements, and expected QoS. \\

    \item \textbf{Split Revision (SR):}
    Implements the logic to re-partition the LFM at different layer boundaries or blocks. This module may use heuristic, rule-based, or learning-based strategies to identify improved splits, respecting constraints such as local privacy boundaries. \\

    \item \textbf{Reconfiguration Broadcast (RB):}
    Propagates new model partitions or sub-partitions to the selected nodes and updates local or remote orchestrators, ensuring future inference requests follow the revised configuration.
\end{enumerate}

Our approach dynamically adapts split inference to fluctuating conditions while maintaining strict QoS and privacy requirements by combining these modules. 
Although \textbf{Kubernetes} already offers node-level telemetry, a pod scheduler, and \texttt{ReplicaSet} roll-outs, it treats the AI model as an \emph{opaque container}. Instead, our \emph{Monitoring\,\&\,CP} module augments the standard metrics stream with \emph{layer-granular} latency, activation size, and privacy tags. The \emph{Adaptive Orchestrator} then optimizes both placement and split boundaries, whereas Kubernetes can only relocate whole pods. Finally, the \emph{Reconfiguration Broadcast} disseminates on-the-fly–generated weight shards and graph rewiring commands—operations that cannot be expressed through a deployment update or \texttt{ConfigMap} patch. Collectively, these extensions lift orchestration from container granularity to \emph{LFM-graph granularity}, which is essential for runtime re-splitting under QoS and privacy constraints.

\ 

\noindent
The next subsections formalize the system model, define constraints, and describe the orchestration workflow for LFMs in detail.

\subsection{Notation and System Model}

We define key terminologies and orchestration concepts that underlie our adaptive split inference framework as follows. \\

\noindent
\textbf{Edge Nodes and Cloud.}
Let $\mathcal{N} = \{1, 2, \ldots, n\}$ denote the set of $n$ edge nodes, and let $c$ refer to a (potential) cloud node (more capacity but increased latency). Each node $j \in \mathcal{N} \cup \{c\}$ has resource capacities for inference at time $t$, captured via capacity profiling (CP) as:
\begin{align}
    \text{CP}(n_j, t) = \{ & \mathrm{CPU}_j(t), \mathrm{GPU}_j(t), \nonumber  \\
    &\mathrm{Mem}_j(t), \mathrm{NetCap}_j(t) \} ,
\end{align}
which vary with concurrent workloads, hardware, and network conditions. \\

\noindent
\textbf{Model Partitioning.}
Consider an LFM \(\mathcal{M}\) segmented into $k$ partitions or layers (e.g., from Transformer or neural networks architectures):
\begin{equation}
    S \;=\; \{S_1, S_2, \dots, S_k\}.
\end{equation}
Typically, $S_1$ handles raw (potentially private) data, and $S_k$ generates the final outputs. Intermediate segments $S_2, \dots, S_{k-1}$ often encompass the bulk of computation (e.g., multi-head self-attention in Transformers). A three-split example $(S_1, S_2, S_3)$ might place $S_1, S_3$ on a local edge node (for privacy where user data is translated into/from vector embeddings) and offload the compute-intensive $S_2$ to a more capable edge or cloud node. Depending on the specific LFM architecture, splits may either be configured as self-contained building blocks (e.g., embeddings, self-attentions) or individual layers (e.g., from deep or convolutional neural networks) \cite{karjee2021split, djuhera2024rsfllmjammingresilientframework}. \\   

\noindent
\textbf{Inference Requests.}
Inference tasks arrive as requests \(\{r_1, r_2, \dots\}\), each with an associated workload \(\mathcal{W}_r\). At a high level, each request utilizes the same model partitions \(\{S_1, \ldots, S_k\}\), but may require separate scheduling decisions depending on QoS constraints or real-time node capacities. We can treat each request as an instance of the partition assignment problem or, if simultaneous requests must be handled, sum over their respective costs when formulating a corresponding objective function. \\

\noindent
\textbf{Decision Variables.}
For computational convenience, we define a binary placement matrix \(\mathbf{x} = [x_{i,j}]\), where \(x_{i,j} = 1\) indicates partition \(S_j\) is assigned to node \(n_i\), and \(x_{c,j} = 1\) indicates assignment to the cloud node \(c\). Each column corresponds to a partition, and each row to a node in \(\mathcal{N}\cup\{c\}\). When multiple requests are considered, either the same \(\mathbf{x}\) can be reused if the system enforces a single partition layout, or a time/index extension can be introduced (e.g., \(\mathbf{x}_{r}\) for each request \(r\)). 

\ 

\noindent
With these concepts and terminologies in place, we may define an appropriate optimization objective as follows. \\

\noindent
\textbf{Objective Function.}
We aim to minimize the high-level cost:
\begin{align}
    \Phi(\mathbf{x}, \mathcal{C}(t))
    \;=\; 
    \alpha\,\mathcal{L}\bigl(\mathbf{x}, \mathcal{C}(t)\bigr)
    &+\;
    \beta\,\mathcal{U}\bigl(\mathbf{x}, \mathcal{C}(t)\bigr)
    \nonumber \\
    &+\;
    \gamma\,\mathcal{P}\bigl(\mathbf{x}, \mathcal{C}(t)\bigr),
    \label{eq:cost_function}
\end{align}
where:
\begin{itemize}
    \item \(\mathcal{L}\) measures inference latency, including data transfer.
    \item \(\mathcal{U}\) captures resource usage imbalance or node overload.
    \item \(\mathcal{P}\) penalizes privacy violations (e.g., placing sensitive partitions on untrusted nodes).
    \item \(\alpha, \beta, \gamma \ge 0\) weight the relative importance of latency, resource usage, and privacy, respectively.
\end{itemize}
Here, \(\mathcal{C}(t)\) encapsulates the system state at time \(t\), including node capacities, network bandwidths, and any QoS or SLA requirements. In scenarios with multiple concurrent requests, \(\Phi\) can be extended to represent the sum or average cost across all active requests. In addition, to ensure valid assignments, we impose the following constraints:

\begin{enumerate}
    \item \textbf{Unique Assignment.}
    Each partition \(S_j\) must be placed on exactly one node:
    \begin{equation}
        \sum_{i \in \mathcal{N}} x_{i,j} \;+\; x_{c,j} = 1,
        \quad
        \forall j \in \{1,\dots,k\}.
    \end{equation} \\
    
    \item \textbf{Capacity Limits.}
    For each node \(n_i \in \mathcal{N}\), the sum of resource loads from its assigned partitions cannot exceed that node’s capacity:
    \begin{equation}
        \sum_{j=1}^k \mathrm{load}\bigl(S_j\bigr)\,x_{i,j} 
        \;\;\le\;\; 
        \mathrm{capacity}(n_i, t).
    \end{equation}
    An analogous constraint applies to the cloud node \(c\) if cloud resources are finite. \\

    \item \textbf{Privacy Constraints.}
    Partitions handling sensitive data (e.g., \(S_1\)) must remain on trusted nodes:
    \begin{equation}
        x_{i,j} = 0, 
        \quad
        \text{if } n_i \notin \mathrm{trustedSet} \;\wedge\; \bigl(S_j\text{ is privacy-critical}\bigr).
    \end{equation}
    
\end{enumerate}

Further, if LFM layer boundaries can be modified (e.g., subdividing $S_2$ into $\{S_{2a}, S_{2b}\}$, as for example in neural network layers, Tranformer embeddings and attentions, etc.), we may treat the set of partitions $S$ itself as part of the optimization. Herewith, we define the split revision as follows. \\

\noindent
\textbf{Split Revision.}
Let \(\Omega\) denote the set of all valid splitting schemes. The orchestrator aims to solve:
\begin{equation}
\min_{\;S \in \Omega,\;\mathbf{x}} 
\;\;
\Phi\!\bigl(\mathbf{x}, S, \mathcal{C}(t)\bigr),
\label{eq:splitrevision}
\end{equation}
to find an optimal split \(S^*\) and assignment \(\mathbf{x}^*\) that minimize the overall cost subject to the constraints above. This allows partitions and assignments to adapt dynamically to shifts in resource availability, privacy requirements, or workload demands, initiated and managed by the adaptive orchestrator.

\subsection{Orchestration Workflow}

Algorithm~\ref{alg:workflow} outlines the main orchestration steps. The workflow begins by deploying a \emph{baseline} partitioning (e.g., $(S_1, S_2, S_3)$) among a set of nodes. The system then continuously monitors resource usage and performance metrics to trigger dynamic adjustments.

\begin{enumerate}
    \item \textbf{Initial Deployment.} Perform a static partitioning of the model based on coarse performance estimates (e.g., place $S_1, S_3$ locally for privacy, and put $S_2$ on a more powerful node or cloud instance~$c$). \\

    \item \textbf{Continuous Monitoring.} The Monitoring \& CP module collects real-time metrics $\mathrm{CP}(n_j, t)$ and calculates an \emph{environment state} $\mathbf{E}(t)$ that captures fluctuations in node utilization, network throughput, or latency. 

    \newpage

    \begin{algorithm}[t]
    \small
    \caption{Adaptive Split Orchestration Workflow}
    \label{alg:workflow}
    \DontPrintSemicolon
    \KwIn{
      (i) Initial partitioning $\{S_1,\dots,S_P\}$, 
      (ii) baseline mapping $d_0$, 
      (iii) monitoring intervals $\Delta t$, 
      (iv) trigger‐threshold vector $\Theta=\{L_{\max},U_{\max},B_{\min},T_{\text{cool}}\}$
    }
    \SetAlgoLined
    
    \textbf{Initialize:} Deploy baseline split $(S_1,\ldots,S_P)$ across nodes as per $d_0$.\\
    Set $t_{\mathrm{last}}\gets -\infty$.\\
    
    \For{\emph{each monitoring cycle} $t \leftarrow 0, \Delta t, 2\Delta t, \ldots$}{
      Collect environment metrics $\mathbf{E}(t)$ via Monitoring \& CP.\\
      $\mathit{reconf}\gets \texttt{ShouldReconfigure}(\mathbf{E}(t),\Theta)$.\\
      \uIf{(\emph{trigger condition is met, e.g., high latency, node overload, etc.}) \textbf{and} $\mathit{reconf}$}{
        Evaluate feasible mappings $\{d^\prime\}$ given current partitions.\\
        Optionally call Model Re-Splitting to produce new partitions $\{S_i^*\}$.\\
        Determine best mapping $\hat{d} = \underset{d^\prime}{\arg\min}\,\mathcal{C}(d^\prime)$.\\
        \If{$\hat{d}\neq d_t$ \textbf{and} $t - t_{\mathrm{last}}\ge T_{\text{cool}}$}{
           Broadcast reconfiguration to all affected nodes via RB.\\
           $t_{\mathrm{last}}\gets t$; \quad $d_{t+\Delta t}\gets \hat{d}$.\\
        }
      }
      Resume inference under current assignment $d_{t+\Delta t}$.\\
    }
    \end{algorithm}
    
    \item \textbf{Adaptive Decisions.}
    Based on the updated system states $\mathcal{C}(t), \mathbf{E}(t)$, the adaptive orchestrator continuously evaluates whether to \emph{keep the current split} (if performance remains within SLA targets), \emph{redistribute sub-splits} (reassigning some partitions $S_j$ from node $n_i$ to $n_{i'}$ by adjusting $\mathbf{x}$ without altering the partition boundaries), or \emph{perform full re-splitting} (to obtain a better partition set $S^*$ via the \emph{SR} module if incremental changes are insufficient or new privacy constraints arise). More formally, 
    \begin{itemize}
        \item The adaptive orchestrator evaluates whether the current partition mapping $d_t$ remains optimal under $\mathbf{E}(t)$. For each request~$r$, the orchestrator checks:
        \begin{equation}
            \mathcal{C}(d_t) \;\overset{?}{\leq}\; \mathcal{C}(d^\prime)
            \quad
            \forall \; \text{feasible } d^\prime.
        \end{equation}
        If a lower-cost (or higher-utility) mapping $d^\prime$ is found, a reconfiguration is triggered.
    
        \item If needed, the SR module modifies the set of partitions $\{S_1,\ldots,S_P\}$ (e.g., subdividing a large block $S_2$ into new split configurations $\{S_{2a}, S_{2b}\}$), i.e.
        \begin{equation}
            \hat{d} \;=\; \underset{d \,\in\, \mathcal{D}(\text{new splits})}{\operatorname{argmin}} \,\mathcal{C}(d),
        \end{equation}
        subject to constraints (e.g., compute, network, privacy).
    \end{itemize} 

    \ 

    \item \textbf{Reconfiguration Broadcast (RB).} Once a decision is made, the \emph{RB} module disseminates the updated assignment $\mathbf{x}^*$ or partition set $S^*$ to relevant nodes, ensuring the new configuration is deployed consistently. \\ 

    \item \textbf{Execution.} Inference resumes with the updated partition assignment $\hat{d}$. The orchestrator continues to monitor performance, forming a feedback loop, allowing the system to adapt further as conditions evolve.

\end{enumerate}




\noindent
\textbf{Additional Trigger Conditions and Decision Logic}.
Table~\ref{tab:thr} summarizes the runtime metrics that feed the
function \texttt{ShouldReconfigure}(E(t),$\Theta$) in Algorithm \ref{alg:workflow}.
The orchestrator invokes a reconfiguration if \emph{any} of the
following holds for a monitoring window of length $\Delta t$:

\begin{enumerate}
  \item \textbf{Latency threshold.}  The exponentially weighted moving average (EWMA) of end-to-end inference latency $\bar{L}(t, \Delta t)$ exceeds $L_{\max}$, i.e. $\bar{L}(t,\Delta t) > L_{\max}$. \\
  
  \item \textbf{Utilisation threshold.}  The maximum GPU or CPU utilization across all nodes exceeds $U_{\max}$, i.e., $\max_{n \in \mathcal{N}} U_n(t) > U_{\max}$. \\ 
  
  \item \textbf{Bandwidth drop.}  The minimum available bandwidth across any active edge link drops below $B_{\min}$, i.e., $\min_{(i,j) \in \mathcal{L}} B_{ij}(t) < B_{\min}$. \\
  
  \item \textbf{Privacy policy violation.} A new inference request is tagged with the identifier \texttt{privacy=high}, but the current partitioning would route raw features through an untrusted node.
\end{enumerate}

\vspace{1ex}
\begin{table}[t]
\centering
\caption{Monitored metrics and default trigger thresholds.}
\label{tab:thr}
\begin{tabular}{lcc}
\toprule
\textbf{Metric} & \textbf{Symbol} & \textbf{Default value} \\
\midrule
EWMA latency                             & $L_{\max}$   & 150 ms \\
GPU/CPU utilisation                      & $U_{\max}$   & 0.85    \\
Available link bandwidth (edge $\to$ edge) & $B_{\min}$   & 50 Mbps \\
Time-to-reconfigure cool-down            & $T_{\text{cool}}$ & 30 s \\
\bottomrule
\end{tabular}
\end{table}

\noindent
If multiple triggers fire simultaneously, the system first attempts \emph{placement} migration. 
If that cannot meet all constraints, the \emph{split-revision} module is invoked. 
Reconfigurations are rate-limited by $T_{\text{cool}}$ to prevent thrashing.
The full control loop of our adaptive orchestration algorithm is illustrated in Figure~\ref{fig:algorithm_workflow}, detailing its monitoring, decision, and reconfiguration stages.

\begin{figure}[htbp]
\centering
\scriptsize   
\begin{tikzpicture}[
  node distance=0.6cm and 0.6cm,
  every node/.style={font=\scriptsize},
  block/.style = {
    rectangle, draw, fill=white!90,
    text width=6em, align=center,
    rounded corners, minimum height=2em
  },
  decision/.style = {
    diamond, draw, fill=white!90,
    text width=4em, align=center, aspect=2
  },
  arrow/.style = {thick, -{Stealth}}
]

  \node[block] (init) {
    Initialize: deploy baseline split\\$d_0$, $t_{\rm last}\gets-\infty$
  };
  \node[block, below=of init] (collect) {
    Collect metrics\\$\mathbf{E}(t)$
  };
  \node[decision, below=of collect] (decide) {
    Should Reconf?\vspace{2pt}\\
    $(\mathbf{E},\Theta)$
  };
  \node[block, below left=of decide,xshift=-1em] (no) {Resume\\under $d_{t+\Delta t}$};
  \node[block, below right=of decide,xshift=1em] (eval) {
    Eval feasible $d'$\\maybe ReSplit
  };
  \node[block, below=of eval] (best) {
    $\hat d\gets\arg\min_{d'}\mathcal C(d')$
  };
  \node[decision, below=of best] (check) {
    Reconf \& Cool-down?
  };
  \node[block, below left=of check,xshift=-1em] (nc) {Resume\\under $d_{t+\Delta t}$};
  \node[block, below right=of check,xshift=1em] (bc) {
    Broadcast\\$d_{t+\Delta t}\gets\hat d$\\$t_{\rm last}\gets t$
  };
  \node[block, below=of bc] (resume) {Resume\\under $d_{t+\Delta t}$};
  \node[block, below=of resume] (loop) {Wait $\Delta t$\\$t\!\leftarrow\!t+\Delta t$};

  \draw[arrow] (init)   -- (collect);
  \draw[arrow] (collect)-- (decide);
  \draw[arrow] (decide) -| node[midway,above,left] {No}  (no);
  \draw[arrow] (no)     |- (loop);
  \draw[arrow] (decide) -| node[midway,above,right]{Yes} (eval);
  \draw[arrow] (eval)   -- (best);
  \draw[arrow] (best)   -- (check);
  \draw[arrow] (check)  -| node[midway,above,left] {No}  (nc);
  \draw[arrow] (nc)     |- (resume);
  \draw[arrow] (check)  -| node[midway,above,right]{Yes} (bc);
  \draw[arrow] (bc)     -- (resume);
  \draw[arrow] (resume) -- (loop);

  \draw[arrow]
    (loop.east)
    -- ++(1cm,0)
    |- (collect.east);

\end{tikzpicture}
\caption{Control-flow diagram of the adaptive split orchestration loop described in Algorithm \ref{alg:workflow}. The orchestrator periodically monitors environment metrics and triggers reconfiguration decisions when QoS thresholds or privacy constraints are violated. Feasible placements are evaluated, and, if no cool-down limit is active, a new mapping is broadcast to all nodes.}
\label{fig:algorithm_workflow}
\end{figure}

\ 

Above outlined system model and orchestration workflow provide possible entry points for optimizations in real-world deployments. In practice, such an orchestration loop can be integrated into existing container platforms (e.g., extending Kubernetes with a custom controller that triggers model re-splitting when monitoring thresholds are exceeded). Partitioning decisions may rely on traditional heuristics (e.g., rule-based or greedy approaches) or adopt learning-based schemes (e.g., reinforcement learning) to continuously refine splitting strategies \cite{optimal_ai_splitting, tuli2022splitplaceaiaugmentedsplitting, lien2024optimum}. Alternatively, Python-based pipelines could invoke layer-partitioning heuristics based on state-of-the-art open-source frameworks, such as Huggingface, and broadcast updates via RESTful APIs. These approaches enable straightforward adoption of adaptive split inference within both on-premise and cloud-based edge deployments.

\subsection{Privacy and Security Considerations}

A core feature behind split inference is the preservation of data privacy by ensuring critical or sensitive operations remain on a trusted device or node. Thus, our framework permits:

\begin{enumerate}
    \item \textbf{Selective Local Execution:}
    Some LFM blocks, especially those close to the input layer, may handle raw personal or private data. By design, these partitions can be configured to remain on the user's device or a trusted edge node (e.g., for compliance with GDPR). Formally, if $S_i$ handles privacy-critical data, we require that
    \begin{equation}
        d_t(i) \;\in\; \mathcal{N}_{\text{trusted}} \quad \forall t,
    \end{equation}
    where $\mathcal{N}_{\text{trusted}} \subseteq \mathcal{N}\cup\{c\}$ is the set of \emph{trusted} nodes. Corresponding LFM splits can be obtained according to the model architecture, compute resources and privacy requirements (e.g., measured as layer depth) \cite{optimal_ai_splitting}. \\ 

    \item \textbf{Secure Communication Channels:}
    Intermediate activations (e.g., outputs from $S_1$ that serve as inputs to $S_2$) can be additionally encrypted and transmitted securely to the next node in the chain. This ensures that eavesdropping or tampering with partial model data (e.g., due to jamming wireless transmissions \cite{djuhera2024rsfllmjammingresilientframework}) is substantially harder. Further, the \emph{RB} component may include additional cryptographic signatures so that only valid reconfiguration commands from the orchestrator are honored. \\

    \item \textbf{Partition Metadata Obfuscation:}
    To further reduce risk, the orchestrator can store only references to partial model weights or encrypted partitions in a registry accessible to each node, such that no single node (other than the one hosting a given partition) stores the raw weights.

\end{enumerate}

\noindent 
Our orchestration framework thus extends standard orchestration platforms but adds specialized components for real-time capacity profiling, model splitting, and reconfiguration in response to varying network and compute conditions. By leveraging partial splits of LFM layers, the framework also inherently supports privacy-preserving inference at the edge, ensuring that sensitive data never leaves a trusted domain.



\newpage

\section{Use Cases}
\label{sec:usecases}

\subsection{Emergency Coordination in Smart Cities}

In a highly connected smart city environment, where autonomous AI agents are responsible for managing infrastructure, monitoring public safety, and responding to critical incidents \cite{saad2024artificialgeneralintelligenceaginative}, large-scale AI inference is crucial for maintaining operational efficiency. During emergency scenarios, such as regional blackouts, cyber-attacks on urban infrastructure, or natural disasters, adaptive split inference ensures real-time decision-making despite fluctuating resource availability.

\

Consider a scenario where a massive earthquake disrupts transportation networks, damages critical infrastructure, and impairs traditional cloud connectivity. Smart city AI agents deployed across distributed MEC nodes can play a pivotal role in orchestrating emergency response through \emph{AI Agents for Autonomous Coordination}, i.e. specialized AI agents trained for disaster response. This can be part of infrastructure monitoring bots, autonomous drones, and emergency service assistants, which rely on continuous, high-throughput AI inference. For instance, these agents need to process high-dimensional multi-modal data, including real-time video, LiDAR scans, and sensor data from IoT devices to provide continuous responses.

\ 

In this environment, \emph{Adaptive Model Deployments} are initially instantiated as foundation model partitions, which are distributed across MEC nodes based on predefined computational capabilities and expected workloads. As infrastructure degradation leads to unstable connectivity and hardware failures, the system dynamically adjusts model partitions across available MEC nodes. For instance, if an AI agent controlling autonomous emergency drones detects a surge in demand for real-time object detection (e.g., identifying survivors in debris), the system triggers split revision \emph{SR} to redistribute workloads efficiently. Consequently, when an MEC node reaches its computational threshold due to a high influx of emergency data streams, the Reconfiguration Broadcast \emph{RB} mechanism ensures that AI agents can offload inference tasks to alternative nodes with idle capacity. The system dynamically revises the model split $S = (S1, S2, S3)$ into a more optimized configuration $S^* = (S^*_1, S^*_2, S^*_3)$, continuously adapting to the dynamic conditions to maintain operational efficiency and robust performance.

\subsection{Industry 4.0 Manufacturing Lines}

Modern manufacturing floors increasingly integrate edge AI for tasks like predictive maintenance, anomaly detection, and quality control, often under tight latency requirements. In such environments, multiple MEC nodes (e.g., private 5G/6G-MEC base stations) may handle continuous data streams from high-speed sensors and robotic arms. When production ramps up unexpectedly, compute workloads spike and nodes near their capacity limits. The orchestrator then responds accordingly by reassigning or re-splitting the inference model across less-loaded nodes, preventing bottlenecks. Privacy-constrained segments, such as those inspecting proprietary designs, remain on trusted hardware, while more generic modules can be offloaded seamlessly to boost throughput.

\subsection{Autonomous Vehicles and Intelligent Transport Systems}

Vehicle-to-infrastructure (V2I) services increasingly rely on advanced AI models for collision avoidance, route planning, and traffic flow optimization. Edge nodes at roadside units (RSUs) offer local compute to complement on-board vehicle processors, reducing latency while offloading computationally heavy layers. In busy urban corridors, traffic sensors and autonomous cars generate significant inference workloads. If congestion surges or a particular RSU becomes overloaded, the orchestrator redistributes model partitions among neighboring edge nodes, ensuring split inference scales effectively. Such real-time adaptivity allows vehicles to maintain continuous, low-latency awareness, meeting strict safety and efficiency standards that are critical in next-generation transport systems.

\section{Expected Results and Quantitative Benefits}

\begin{table*}[t]
\centering
\scriptsize
\caption{Quantitative comparison of static and adaptive split inference across key performance dimensions in edge environments. Results are sourced from prior studies and testbed evaluations of LLM deployments over 5G-MEC infrastructures \cite{itu2020imt, sarah2023resource, xu2021survey, karjee2021energy, zhang2025amp4ecadaptivemodelpartitioning, tuli2022splitplaceaiaugmentedsplitting, chen2024adaptive}.}
\label{tab:perf_compare}
\begin{tabularx}{\linewidth}{lXX}
\toprule
\textbf{Performance Dimension} 
  & \textbf{Static Split Inference} 
  & \textbf{Adaptive Split Inference} \\
\midrule
\textbf{Latency (End-to-End)} 
  & High variance; $\sim$~500-1000\,ms typical in 5G-MEC scenarios. Cannot guarantee low latency under network fluctuation (may spike beyond 1\,s). 
  & Low and stable; $\sim$~100-300\,ms under same conditions. Dynamically maintains latency below target (often 50\%+ faster than static). \\[4pt]

\addlinespace

\textbf{Throughput} 
  & Limited by weakest link (device or network); lower overall. $<1\,\mathrm{req/s}$ on a baseline edge device alone. 
  & Higher via parallelism and load balancing; utilizes multiple nodes. Achieved $>5\,\mathrm{req/s}$ in adaptive multi-node setup ($\geq5$× higher). \\[4pt]

\addlinespace

\textbf{Resource Utilization} 
  & $\sim$~50-60\% of available edge / cloud resources used. Static partition leads to idle resources on one side and potential overload on the other. 
  & $\sim$ 80-95\% resource utilization. Orchestrator keeps both edge and cloud busy, scaling across nodes; more efficient use of CPU, GPU, and bandwidth. \\[4pt]

\addlinespace

\textbf{QoS Compliance (SLA HitRate)} 
  & Often poor under variability. QoS deadlines met in $\sim$ 60-70\% of cases; frequent SLA violations when conditions deviate from design point. 
  & Near-guaranteed QoS even as conditions change. $\sim$ 95-99\% of requests meet latency/SLA targets due to real-time adaptation (much fewer deadline misses). \\[4pt]

\addlinespace

\textbf{Reliability \& Failover} 
  & Rigid deployment; single-point bottlenecks. $\sim$ 5-10 inference failures or timeouts per hour observed under load/network issues. 
  & Resilient, with dynamic re-routing and re-partitioning. Downtime incidents reduced to $\sim$ 0-2 per hour. Maintains service continuity by avoiding overloads. \\[4pt]

\addlinespace

\textbf{Privacy Adherence} 
  & Moderate. Fixed layer offload may send sensitive features off device. No ability to alter behavior for sensitive data—potential compliance issues. 
  & High. Can execute sensitive layers locally and limit data exposure. Adaptive policy balances performance with data confidentiality (privacy preserved without sacrificing QoS). \\
\bottomrule
\end{tabularx}
\end{table*}

Table~\ref{tab:perf_compare} presents a comparative analysis derived from prior studies, experimental testbeds, and empirical estimates for a representative 5G-MEC-enabled edge environment serving text-generation LLMs ranging from 7B to 13B parameters. For further technical depth and context, we refer  to the comprehensive evaluations in \cite{itu2020imt, sarah2023resource, xu2021survey, karjee2021energy, zhang2025amp4ecadaptivemodelpartitioning, tuli2022splitplaceaiaugmentedsplitting, chen2024adaptive}.

\

In particular, it becomes evident that adaptive split inference offers significant improvements across multiple operational dimensions. First, it consistently achieves lower and more stable end-to-end latency, typically maintaining inference response times within 100–300ms, even under fluctuating 5G-MEC conditions, whereas static configurations frequently exceed 500ms with high variance. This translates to a latency reduction of over 50\% in representative scenarios. Second, throughput is substantially improved: adaptive orchestration enables parallel execution across distributed nodes, reaching over 5 req/s compared to sub-1 req/s throughput observed in static, single-device deployments.

\ 

Resource utilization also benefits considerably. Static splits often lead to underutilized compute on one end (e.g., idle cloud GPUs) and overloading on the other, capping effective usage at 50–60\%. In contrast, adaptive orchestration dynamically balances workloads, yielding sustained utilization rates of 80–95\% across edge and cloud components. This directly contributes to higher QoS adherence: while static inference often fails to meet SLA constraints during network or load fluctuations, achieving target latency in only 60–70\% of cases, adaptive inference maintains compliance for 95–99\% of requests through real-time partition and placement adjustments.

\ 

Further, reliability under stress is markedly improved. Static deployments are prone to bottlenecks and failure cascades, with up to 10 inference errors or timeouts per hour observed under load. Adaptive strategies mitigate this via failover-aware routing and reconfiguration, reducing service interruptions to near-zero levels. In addition, privacy adherence is strengthened. While static splits may expose sensitive intermediate representations to untrusted infrastructure which can be targeted by malicious attackers, the adaptive approach can selectively retain privacy-critical layers on-device, enabling compliance with confidentiality requirements without degrading system performance.

\ 

\begin{figure}[t]
  \centering
  \includegraphics[width=0.8\linewidth]{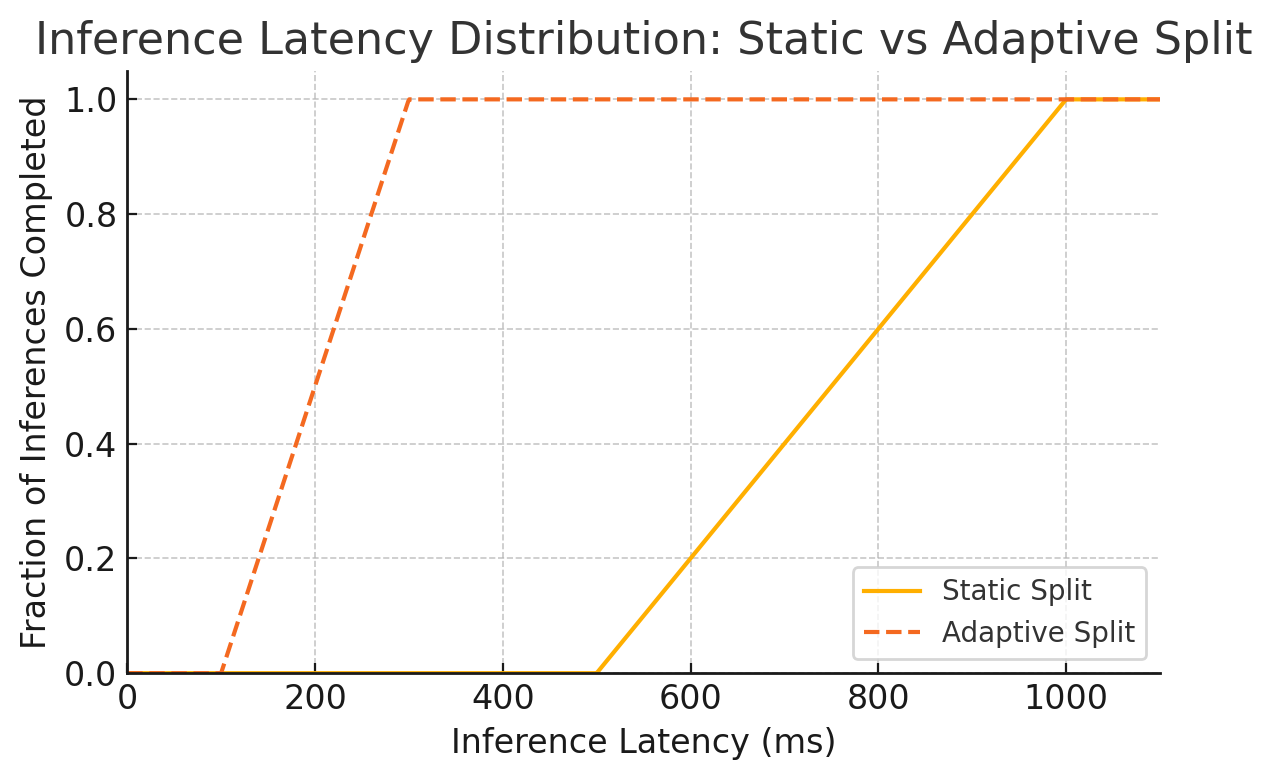}
  \caption{CDF of end-to-end inference latency for static (solid) vs.\ adaptive
  (dashed) split inference in a 5G-MEC scenario. 95\,\% of adaptive
  requests finish within 300 ms, while static requests may take up to 1s \cite{zhang2025amp4ecadaptivemodelpartitioning, tuli2022splitplaceaiaugmentedsplitting, EdgeShard, mudvari2024adaptivecompressionawaresplitlearning}.}
  \label{fig:cdf_latency}
\end{figure}

Table \ref{tab:static_vs_adaptive} summarizes the discussed median figures reported
across four public 5G–MEC studies \cite{zhang2025amp4ecadaptivemodelpartitioning, tuli2022splitplaceaiaugmentedsplitting, EdgeShard, mudvari2024adaptivecompressionawaresplitlearning} and corresponding latency distributions are visualized in Figure~\ref{fig:cdf_latency} (static vs.\ adaptive CDF).  The adaptive curve reaches the 95\,\% completion mark below $300$\,ms, while the static curve stretches beyond $1$s, highlighting the practical QoS benefit of adaptive orchestration at runtime.

\begin{table}[t]
\centering
\caption{Static vs. adaptive split inference for text-generation LLMs in 
typical 5G‐MEC deployments (median values across the studies in \cite{zhang2025amp4ecadaptivemodelpartitioning, tuli2022splitplaceaiaugmentedsplitting, EdgeShard, mudvari2024adaptivecompressionawaresplitlearning}).}
\label{tab:static_vs_adaptive}
\setlength{\tabcolsep}{5pt}
\begin{tabular}{@{}lcc@{}}
\toprule
\textbf{Metric} & \textbf{Static Split} & \textbf{Adaptive Split} \\
\midrule
End-to-end latency                & 500–1000 ms     & 100–300 ms   \\
Throughput (requests s$^{-1}$)    & $\sim$1         & $\sim$5      \\
GPU/CPU utilisation               & 50–60\,\%       & 80–95\,\%    \\
SLA hit-rate (400 ms budget)      & 60–70\,\%       & 95–99\,\%    \\
Downtime incidents (per h)        & 5–10            & 0–2          \\
Privacy compliance                & Moderate        & High         \\
\bottomrule
\end{tabular}
\end{table}

\ 

Across all evaluative dimensions (latency, throughput, utilisation, QoS robustness, and privacy), \textit{adaptive split inference} consistently outperforms any static configuration. The small overhead of monitoring ($\le$10 ms per cycle) and graph rewiring is amortised by hundreds of ms saved per request, yielding a net performance gain an order of magnitude larger than the orchestration cost. Consequently, adaptive split inference emerges as the decisive design choice for LLM deployment in heterogeneous, bandwidth-variable 5G/6G edge networks.

\section{Conclusions}
\label{sec:conclusions}

This study has introduced an adaptive split inference orchestration framework designed to dynamically manage LFM partitions across heterogeneous edge nodes, addressing the inefficiencies inherent in static split inference methodologies. Our framework establishes a foundation for real-time, QoS-aware, and privacy-preserving AI inference in edge computing environments, which is particularly crucial for latency-sensitive and resource-constrained applications. The proposed approach optimizes performance, enhances resource efficiency, and fortifies privacy preservation by leveraging real-time monitoring, workload redistribution, and dynamic reconfiguration. In contrast to existing orchestration frameworks, we do not treat AI models as opaque containers, but instead support runtime model re-partitioning and privacy-aware layer placement, capabilities essential for deploying LFMs in dynamic edge environments.
Our proposed orchestration model thus aligns with emerging objectives in the development of 6G networks by enabling intelligent, distributed AI processing at the edge. Its modular architecture facilitates integration with existing edge orchestrators while maintaining extensibility for future AI-driven optimizations. 

\ 

To validate the practical relevance of our framework, we provided a comparative analysis grounded in prior studies, testbed results, and empirical evaluations. These results demonstrate consistent gains across latency, throughput, resource utilization, reliability, and privacy adherence, reinforcing the value of adaptive orchestration for real-world edge inference given the complexity of future foundation model deployments. 

\

Nevertheless, several research directions emerge to further advance adaptive inference orchestration. Proposed and necessary items for a research agenda on future work include:

\begin{enumerate}

    \item Investigating the deeper integration of AI-driven decision-making mechanisms, including reinforcement learning-based optimizations, to enhance inference orchestration. \\

    \item Developing advanced privacy-preserving techniques, such as secure multi-party computation and homomorphic encryption, to ensure robust data security in distributed inference environments. \\

    \item Designing adaptive network-aware partitioning strategies that dynamically adjust inference workload distribution based on real-time network conditions to optimize resource utilization and latency minimization. \\

    \item Establishing standardized benchmarks and datasets for evaluating the performance of split inference frameworks in edge computing environments.

\end{enumerate}

Beyond technical extensions, domain-specific deployments in autonomous systems, healthcare, and industrial automation present high-impact opportunities for practical validation. Likewise, examining the role of edge-driven inference in reducing energy consumption and enabling sustainable AI operations across smart cities and IoT ecosystems opens new interdisciplinary research avenues. Addressing these challenges will be essential to realizing scalable, trustworthy, and privacy-preserving AIaaS architectures in the forthcoming 6G edge landscape.



\section*{Acknowledgment}

This paper provides a scholarly description of the published Patent US20250071069\-A1 \cite{Djuhere2025patent}, along with references to published Patent US18/449811 \cite{Djuhere2024patent}, which have been conceived and authored while the authors were employed at International Business Machines (IBM). IBM is the current assignee of both patent applications.


\bibliographystyle{IEEEtran} 
\bibliography{mybib}

\begin{thebibliography}{10}
\providecommand{\url}[1]{#1}
\csname url@samestyle\endcsname
\providecommand{\newblock}{\relax}
\providecommand{\bibinfo}[2]{#2}
\providecommand{\BIBentrySTDinterwordspacing}{\spaceskip=0pt\relax}
\providecommand{\BIBentryALTinterwordstretchfactor}{4}
\providecommand{\BIBentryALTinterwordspacing}{\spaceskip=\fontdimen2\font plus
\BIBentryALTinterwordstretchfactor\fontdimen3\font minus \fontdimen4\font\relax}
\providecommand{\BIBforeignlanguage}[2]{{%
\expandafter\ifx\csname l@#1\endcsname\relax
\typeout{** WARNING: IEEEtran.bst: No hyphenation pattern has been}%
\typeout{** loaded for the language `#1'. Using the pattern for}%
\typeout{** the default language instead.}%
\else
\language=\csname l@#1\endcsname
\fi
#2}}
\providecommand{\BIBdecl}{\relax}
\BIBdecl

\bibitem{zhou2024survey}
Z.~Zhou, X.~Ning, K.~Hong, T.~Fu, J.~Xu, S.~Li, Y.~Lou, L.~Wang, Z.~Yuan, X.~Li \emph{et~al.}, ``{A Survey on Efficient Inference for Large Language Models},'' \emph{arXiv preprint arXiv:2404.14294}, 2024.

\bibitem{vaswani2017attention}
A.~Vaswani, N.~Shazeer, N.~Parmar, J.~Uszkoreit, L.~Jones, A.~N. Gomez, {\L}.~Kaiser, and I.~Polosukhin, ``{Attention is All You Need},'' \emph{Advances in neural information processing systems}, vol.~30, 2017.

\bibitem{grattafiori2024llama}
A.~Grattafiori, A.~Dubey, A.~Jauhri, A.~Pandey, A.~Kadian, A.~Al-Dahle, A.~Letman, A.~Mathur, A.~Schelten, A.~Vaughan \emph{et~al.}, ``{The Llama 3 Herd of Models},'' \emph{arXiv preprint arXiv:2407.21783}, 2024.

\bibitem{yang2025qwen3}
A.~Yang, A.~Li, B.~Yang, B.~Zhang, B.~Hui, B.~Zheng, B.~Yu, C.~Gao, C.~Huang, C.~Lv \emph{et~al.}, ``{Qwen3 Technical Report},'' \emph{arXiv preprint arXiv:2505.09388}, 2025.

\bibitem{zhang2024beyond}
X.~Zhang, J.~Nie, Y.~Huang, G.~Xie, Z.~Xiong, J.~Liu, D.~Niyato, and X.~S. Shen, ``{Beyond the Cloud: Edge Inference for Generative Large Language Models in Wireless Networks},'' \emph{IEEE Transactions on Wireless Communications}, 2024.

\bibitem{karjee2022split}
J.~Karjee, P.~{Naik S}, K.~Anand, and V.~N. Bhargav, ``{Split computing: DNN Inference Partition with Load Balancing in IoT-Edge Platform for Beyond 5G},'' \emph{Measurement: Sensors}, vol.~23, p. 100409, 2022.

\bibitem{karjee2021split}
J.~Karjee, K.~Anand, V.~N. Bhargav, P.~S. Naik, R.~B.~V. Dabbiru, and N.~Srinidhi, ``{Split Computing: Dynamic Partitioning and Reliable Communications in IoT-Edge for 6G Vision},'' in \emph{2021 8th International Conference on Future Internet of Things and Cloud (FiCloud)}.\hskip 1em plus 0.5em minus 0.4em\relax IEEE, 2021, pp. 233--240.

\bibitem{GDPR2016a}
\BIBentryALTinterwordspacing
{European Parliament} and {Council of the European Union}. Regulation ({EU}) 2016/679 of the {European} {Parliament} and of the {Council}. [Online]. Available: \url{https://data.europa.eu/eli/reg/2016/679/oj}
\BIBentrySTDinterwordspacing

\bibitem{struye2020towards}
J.~Struye, F.~Lemic, and J.~Famaey, ``{Towards Ultra-Low-Latency mmWave Wi-Fi for Multi-User Interactive Virtual Reality},'' in \emph{GLOBECOM 2020-2020 IEEE Global Communications Conference}.\hskip 1em plus 0.5em minus 0.4em\relax IEEE, 2020, pp. 1--6.

\bibitem{lau2023hybrid}
I.~Lau, S.~Ekpo, M.~Zafar, M.~Ijaz, and A.~Gibson, ``{Hybrid mmWave-Li-Fi 5G Architecture for Reconfigurable Variable Latency and Data Rate Communications},'' \emph{IEEE Access}, vol.~11, pp. 42\,850--42\,861, 2023.

\bibitem{vasireddy2023kubernetes}
I.~Vasireddy, G.~Ramya, and P.~Kandi, ``{Kubernetes and Docker Load Balancing: State-of-the-Art Techniques and Challenges},'' \emph{International Journal of Innovative Research in Engineering and Management}, vol.~10, no.~6, pp. 49--54, 2023.

\bibitem{djuhera2024rsfllmjammingresilientframework}
\BIBentryALTinterwordspacing
A.~Djuhera, V.~C. Andrei, X.~Li, U.~J. Mönich, H.~Boche, and W.~Saad, ``{R-SFLLM: Jamming Resilient Framework for Split Federated Learning with Large Language Models},'' 2024. [Online]. Available: \url{https://arxiv.org/abs/2407.11654}
\BIBentrySTDinterwordspacing

\bibitem{jiang2024federated}
W.~Jiang, H.~Han, Y.~Zhang, and J.~Mu, ``{Federated Split Learning for Sequential Data in Satellite-Terrestrial Integrated Networks},'' \emph{Information Fusion}, vol. 103, p. 102141, 2024.

\bibitem{xia2025sfml}
J.~Xia, M.~Wu, and P.~Li, ``{SFML: A Personalized, Efficient, and Privacy-Preserving Collaborative Traffic Classification Architecture Based on Split Learning and Mutual Learning},'' \emph{Future Generation Computer Systems}, vol. 162, p. 107487, 2025.

\bibitem{xu2021privacy}
R.~Xu, N.~Baracaldo, and J.~Joshi, ``{Privacy-Preserving Machine Learning: Methods, Challenges and Directions},'' \emph{arXiv preprint arXiv:2108.04417}, 2021.

\bibitem{letaief2021edge}
K.~B. Letaief, Y.~Shi, J.~Lu, and J.~Lu, ``{Edge Artificial Intelligence for 6G: Vision, Enabling Technologies, and Applications},'' \emph{IEEE journal on selected areas in communications}, vol.~40, no.~1, pp. 5--36, 2021.

\bibitem{saad2024artificialgeneralintelligenceaginative}
\BIBentryALTinterwordspacing
W.~Saad, O.~Hashash, C.~K. Thomas, C.~Chaccour, M.~Debbah, N.~Mandayam, and Z.~Han, ``{Artificial General Intelligence (AGI)-Native Wireless Systems: A Journey Beyond 6G},'' 2024. [Online]. Available: \url{https://arxiv.org/abs/2405.02336}
\BIBentrySTDinterwordspacing

\bibitem{llms_telecom}
H.~Zhou, C.~Hu, Y.~Yuan, Y.~Cui, Y.~Jin, C.~Chen, H.~Wu, D.~Yuan, L.~Jiang, D.~Wu, X.~Liu, C.~Zhang, X.~Wang, and J.~Liu, ``{Large Language Model (LLM) for Telecommunications: A Comprehensive Survey on Principles, Key Techniques, and Opportunities},'' \emph{IEEE Communications Surveys \& Tutorials}, pp. 1--1, 2024.

\bibitem{li2024llm}
B.~Li, Y.~Jiang, V.~Gadepally, and D.~Tiwari, ``{LLM Inference Serving: Survey of Recent Advances and Opportunities},'' \emph{arXiv preprint arXiv:2407.12391}, 2024.

\bibitem{yao2024survey}
Y.~Yao, J.~Duan, K.~Xu, Y.~Cai, Z.~Sun, and Y.~Zhang, ``{A Survey on Large Language Model (LLM) Security and Privacy: The Good, the Bad, and the Ugly},'' \emph{High-Confidence Computing}, p. 100211, 2024.

\bibitem{li2019edge}
E.~Li, L.~Zeng, Z.~Zhou, and X.~Chen, ``{Edge AI: On-Demand Accelerating Deep Neural Network Inference via Edge Computing},'' \emph{IEEE transactions on wireless communications}, vol.~19, no.~1, pp. 447--457, 2019.

\bibitem{abderrahime2020mec}
A.~Filali, A.~Abouaomar, S.~Cherkaoui, A.~Kobbane, and M.~Guizani, ``{Multi-Access Edge Computing: A Survey},'' \emph{IEEE Access}, vol.~8, pp. 197\,017--197\,046, 2020.

\bibitem{lin2023pushing}
Z.~Lin, G.~Qu, Q.~Chen, X.~Chen, Z.~Chen, and K.~Huang, ``{Pushing Large Language Models to the 6G Edge: Vision, Challenges, and Opportunities},'' \emph{arXiv preprint arXiv:2309.16739}, 2023.

\bibitem{fettweis20226g}
G.~P. Fettweis and H.~Boche, ``{On 6G and Trustworthiness},'' \emph{Communications of the ACM}, vol.~65, no.~4, pp. 48--49, 2022.

\bibitem{camelo2022requirements}
M.~Camelo, L.~Cominardi, M.~Gramaglia, M.~Fiore, A.~Garcia-Saavedra, L.~Fuentes, D.~De~Vleeschauwer, P.~Soto-Arenas, N.~Slamnik-Krijestorac, J.~Ballesteros \emph{et~al.}, ``{Requirements and Specifications for the Orchestration of Network Intelligence in 6G},'' in \emph{2022 IEEE 19th Annual Consumer Communications \& Networking Conference (CCNC)}.\hskip 1em plus 0.5em minus 0.4em\relax IEEE, 2022, pp. 1--9.

\bibitem{zeb2023toward}
S.~Zeb, M.~A. Rathore, S.~A. Hassan, S.~Raza, K.~Dev, and G.~Fortino, ``{Toward AI-Enabled NextG Networks with Edge Intelligence-Assisted Microservice Orchestration},'' \emph{IEEE Wireless Communications}, vol.~30, no.~3, pp. 148--156, 2023.

\bibitem{chen2021enhancing}
W.~Chen, Y.~Zhu, J.~Liu, and Y.~Chen, ``{Enhancing Mobile Edge Computing with Efficient Load Balancing using Load Estimation in Ultra-Dense Network},'' \emph{Sensors}, vol.~21, no.~9, p. 3135, 2021.

\bibitem{chen2024adaptive}
\BIBentryALTinterwordspacing
Y.~Chen, R.~Li, X.~Yu, Z.~Zhao, and H.~Zhang, ``{Adaptive Layer Splitting for Wireless LLM Inference in Edge Computing: A Model-Based Reinforcement Learning Approach},'' 2024. [Online]. Available: \url{https://arxiv.org/abs/2406.02616}
\BIBentrySTDinterwordspacing

\bibitem{carrion2022kubernetes}
C.~Carri{\'o}n, ``{Kubernetes Scheduling: Taxonomy, Ongoing Issues and Challenges},'' \emph{ACM Computing Surveys}, vol.~55, no.~7, pp. 1--37, 2022.

\bibitem{EdgeShard}
M.~Zhang, X.~Shen, J.~Cao, Z.~Cui, and S.~Jiang, ``{EdgeShard: Efficient LLM Inference via Collaborative Edge Computing},'' \emph{IEEE Internet of Things Journal}, pp. 1--1, 2024.

\bibitem{duan2024efficienttraininglargelanguage}
\BIBentryALTinterwordspacing
J.~Duan, S.~Zhang, Z.~Wang, L.~Jiang, W.~Qu, Q.~Hu, G.~Wang, Q.~Weng, H.~Yan, X.~Zhang, X.~Qiu, D.~Lin, Y.~Wen, X.~Jin, T.~Zhang, and P.~Sun, ``{Efficient Training of Large Language Models on Distributed Infrastructures: A Survey},'' 2024. [Online]. Available: \url{https://arxiv.org/abs/2407.20018}
\BIBentrySTDinterwordspacing

\bibitem{lang2024comprehensive}
J.~Lang, Z.~Guo, and S.~Huang, ``{A Comprehensive Study on Quantization Techniques for Large Language Models},'' in \emph{2024 4th International Conference on Artificial Intelligence, Robotics, and Communication (ICAIRC)}.\hskip 1em plus 0.5em minus 0.4em\relax IEEE, 2024, pp. 224--231.

\bibitem{yao2024federated}
Y.~Yao, J.~Zhang, J.~Wu, C.~Huang, Y.~Xia, T.~Yu, R.~Zhang, S.~Kim, R.~Rossi, A.~Li \emph{et~al.}, ``{Federated Large Language Models: Current Progress and Future Directions},'' \emph{arXiv preprint arXiv:2409.15723}, 2024.

\bibitem{alecio1}
A.~P.~D. Binotto, C.~E. Pereira, A.~Kuijper, A.~Stork, and D.~W. Fellner, ``{An Effective Dynamic Scheduling Runtime and Tuning System for Heterogeneous Multi and Many-Core Desktop Platforms},'' in \emph{2011 IEEE International Conference on High Performance Computing and Communications}, 2011, pp. 78--85.

\bibitem{alecio2}
A.~P.~D. Binotto, M.~A. Wehrmeister, A.~Kuijper, and C.~E. Pereira, ``{Sm@rtConfig: A Context-Aware Runtime and Tuning System using an Aspect-Oriented Approach for Data Intensive Engineering Applications},'' \emph{Control Engineering Practice}, vol.~21, no.~2, pp. 204--217, 2013.

\bibitem{mohammed2020distributed}
T.~Mohammed, C.~Joe-Wong, R.~Babbar, and M.~D. Francesco, ``{Distributed Inference Acceleration with Adaptive DNN Partitioning and Offloading},'' in \emph{IEEE INFOCOM 2020 - IEEE Conference on Computer Communications}, 2020, pp. 854--863.

\bibitem{timor2025distributed}
\BIBentryALTinterwordspacing
N.~Timor, J.~Mamou, D.~Korat, M.~Berchansky, O.~Pereg, M.~Wasserblat, T.~Galanti, M.~Gordon, and D.~Harel, ``{Distributed Speculative Inference (DSI): Speculation Parallelism for Provably Faster Lossless Language Model Inference},'' 2025. [Online]. Available: \url{https://arxiv.org/abs/2405.14105}
\BIBentrySTDinterwordspacing

\bibitem{hudson2024qos}
N.~Hudson, H.~Khamfroush, M.~Baughman, D.~E. Lucani, K.~Chard, and I.~Foster, ``{QoS-Aware Edge AI Placement and Scheduling with Multiple Implementations in FaaS-Based Edge Computing},'' \emph{Future Generation Computer Systems}, vol. 157, pp. 250--263, 2024.

\bibitem{thantharate2023adaptive6g}
A.~Thantharate and C.~Beard, ``{ADAPTIVE6G: Adaptive Resource Management for Network Slicing Architectures in Current 5G and Future 6G Systems},'' \emph{Journal of Network and Systems Management}, vol.~31, no.~1, p.~9, 2023.

\bibitem{zhou2019distributing}
L.~Zhou, H.~Wen, R.~Teodorescu, and D.~H. Du, ``{Distributing Deep Neural Networks with Containerized Partitions at the Edge},'' in \emph{2nd USENIX Workshop on Hot Topics in Edge Computing (HotEdge 19)}, 2019.

\bibitem{li2019learning}
Y.~Li, X.~Wang, X.~Gan, H.~Jin, L.~Fu, and X.~Wang, ``{Learning-Aided Computation Offloading for Trusted Collaborative Mobile Edge Computing},'' \emph{IEEE Transactions on Mobile Computing}, vol.~19, no.~12, pp. 2833--2849, 2019.

\bibitem{optimal_ai_splitting}
X.~Li and S.~Bi, ``{Optimal AI Model Splitting and Resource Allocation for Device-Edge Co-Inference in Multi-User Wireless Sensing Systems},'' \emph{IEEE Transactions on Wireless Communications}, vol.~23, no.~9, pp. 11\,094--11\,108, 2024.

\bibitem{tuli2022splitplaceaiaugmentedsplitting}
\BIBentryALTinterwordspacing
S.~Tuli, G.~Casale, and N.~R. Jennings, ``{SplitPlace: AI Augmented Splitting and Placement of Large-Scale Neural Networks in Mobile Edge Environments},'' 2022. [Online]. Available: \url{https://arxiv.org/abs/2205.10635}
\BIBentrySTDinterwordspacing

\bibitem{lien2024optimum}
S.-Y. Lien, C.-H. Yeh, and D.-J. Deng, ``{Optimum Splitting Computing for DNN Training Through Next Generation Smart Networks: A Multi-Tier Deep Reinforcement Learning Approach},'' \emph{Wireless Networks}, vol.~30, no.~3, pp. 1737--1751, 2024.

\bibitem{itu2020imt}
\BIBentryALTinterwordspacing
{International Telecommunication Union (ITU)}, ``{IMT-2020 (5G) Standard: Minimum Requirements Related to Technical Performance for IMT-2020 Radio Interface(s)},'' International Telecommunication Union, Tech. Rep., 2020, online; accessed 8 March 2025. [Online]. Available: \url{https://www.itu.int/pub/R-REP-M.2410-2020}
\BIBentrySTDinterwordspacing

\bibitem{sarah2023resource}
A.~Sarah, G.~Nencioni, and M.~M.~I. Khan, ``{Resource Allocation in Multi-Access Edge Computing for 5G-and-Beyond Networks},'' \emph{Computer Networks}, vol. 227, p. 109720, 2023.

\bibitem{xu2021survey}
Y.~Xu, G.~Gui, H.~Gacanin, and F.~Adachi, ``{A Survey on Resource Allocation for 5G Heterogeneous Networks: Current Research, Future Trends, and Challenges},'' \emph{IEEE Communications Surveys \& Tutorials}, vol.~23, no.~2, pp. 668--695, 2021.

\bibitem{karjee2021energy}
J.~Karjee, S.~P. Naik, and N.~Srinidhi, ``{Energy Profiling based Load-Balancing Approach in IoT-Edge for Split Computing},'' in \emph{2021 IEEE 18th India Council International Conference (INDICON)}.\hskip 1em plus 0.5em minus 0.4em\relax IEEE, 2021, pp. 1--6.

\bibitem{zhang2025amp4ecadaptivemodelpartitioning}
\BIBentryALTinterwordspacing
G.~Zhang, W.~Guo, Z.~Tan, and H.~Jiang, ``{AMP4EC: Adaptive Model Partitioning Framework for Efficient Deep Learning Inference in Edge Computing Environments},'' 2025. [Online]. Available: \url{https://arxiv.org/abs/2504.00407}
\BIBentrySTDinterwordspacing

\bibitem{mudvari2024adaptivecompressionawaresplitlearning}
\BIBentryALTinterwordspacing
A.~Mudvari, A.~Vainio, I.~Ofeidis, S.~Tarkoma, and L.~Tassiulas, ``{Adaptive Compression-Aware Split Learning and Inference for Enhanced Network Efficiency},'' 2024. [Online]. Available: \url{https://arxiv.org/abs/2311.05739}
\BIBentrySTDinterwordspacing

\bibitem{Djuhere2025patent}
A.~Djuhera, A.~P.~D. Binotto, F.~L. Koch, and R.~High, ``{Orchestration of Workloads Involving an AI Model},'' Patent US20\,250\,071\,069A1, February 27, 2025, patent application filed on October 10, 2023, pending approval.

\bibitem{Djuhere2024patent}
A.~Djuhera, A.~P.~D. Binotto, F.~L. Koch, and N.~Baracaldo~Angel, ``{Distributed Execution of an Aritificial Intelligence Model},'' Patent US 18/449\,811, December 19, 2024.

\end{thebibliography}





\end{document}